\documentclass[preprint]{vgtc}               
\ifpdf
  \pdfoutput=1\relax                   
  \usepackage{graphicx}                
  \DeclareGraphicsExtensions{.pdf,.png,.jpg,.jpeg} 
\else
  \usepackage{graphicx}                
  \DeclareGraphicsExtensions{.eps}     
\fi%

\graphicspath{{figures/}{pictures/}{images/}{./}}
\usepackage{microtype}
\PassOptionsToPackage{warn}{textcomp}
\usepackage{textcomp}
\usepackage{mathptmx}
\usepackage{times}

\usepackage{cite}
\usepackage{tabu}
\usepackage{booktabs}
\usepackage{epsfig}
\usepackage{ellipsis}
\usepackage{epigraph}
\usepackage{subcaption}
\usepackage{multirow}
\usepackage{multicol}
\usepackage{tabularx}
\usepackage{amssymb}
\usepackage{amsmath}
\usepackage[hidelinks]{hyperref}
\usepackage{xcolor}
\usepackage{enumitem}

\definecolor{snsblue}{HTML}{0173b2}
\definecolor{snsorange}{HTML}{de8f05}

\onlineid{0}

\vgtccategory{Research}
\vgtcpapertype{please specify}

\title{Expectation Versus Reality:\\The Failed Evaluation of a Mixed-Initiative Visualization System}

\author{Sunwoo Ha\thanks{Ha is with Washington University in St. Louis: sha@wustl.edu} , Adam Kern\thanks{Kern is with MIT Lincoln Laboratory: adamnatkern@gmail.com} , Melanie Bancilhon\thanks{Bancilhon is with Washington University in St. Louis: mbancilhon@wustl.edu} , and Alvitta Ottley  \thanks{Ottley is with Washington University in St. Louis: alvitta@wustl.edu}
}

\abstract{
Our research aimed to present the design and evaluation of a mixed-initiative system that aids the user in handling complex datasets and dense visualization systems. We attempted to demonstrate this system with two trials of an online between-groups, two-by-two study, measuring the effects of this mixed-initiative system on user interactions and system usability. However, due to flaws in the interface design and the expectations that we put on users, we were unable to show that the adaptive system had an impact on user interactions or system usability. In this paper, we discuss the unexpected findings that we found from our ``failed" experiments and examine how we can learn from our failures to improve further research.
} 

\teaser{
  \centering
    \includegraphics[width=\textwidth]{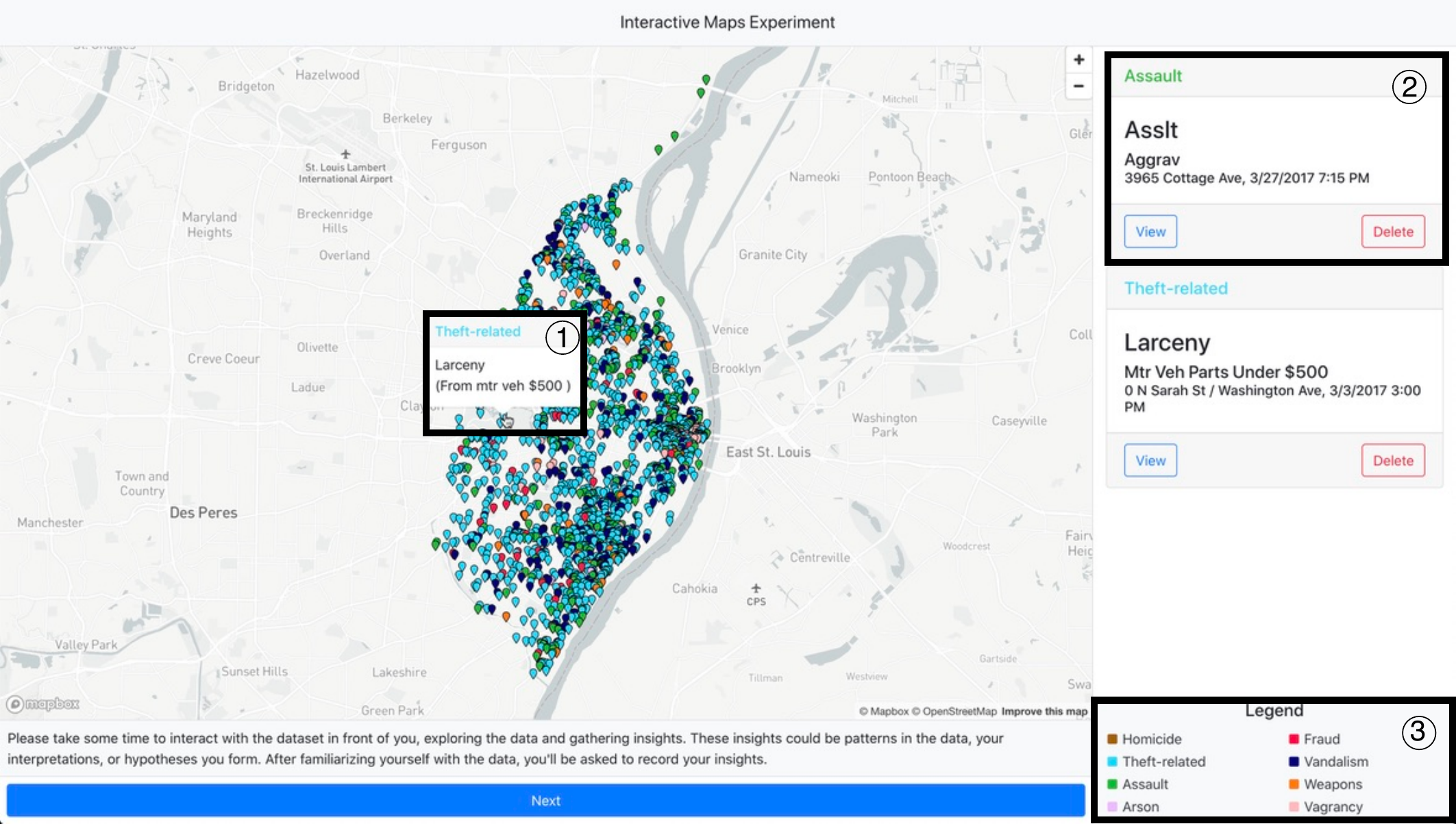}
    \caption{An overview of the system with visualization components labelled. Component 1 corresponds to the hover tooltip; component 2 corresponds to the information card; component 3 corresponds to the label and filter.}\label{fig:exp_main}
}

\begin{document}

\firstsection{Introduction}\label{cpt:intro}

\maketitle

Evaluation in visualization studies show how visual aides can support analysts, researchers, and all who must interact with data of all types in understanding, synthesizing, and communicating that data. Historically, this has been accomplished using controlled, in-person laboratory experiments following practices from the psychology and broader HCI community. With calls for more ecologically valid evaluations that examine actions from a more representative user study population, there has been a relatively recent tendency to collect data via crowdsourced platforms in place of an undergraduate student population.  

Researchers have successfully replicated pioneering studies using platforms such as Amazon Mechanical Turk~\cite{Amazon2020}, and have produced more generalizable findings that involve a more extensive and diverse population (see ~\cite{borgo2018information} for a comprehensive survey of the prior work). However, there are a few caveats. Mechanical Turk data is notoriously noisy, and researchers typically need to use techniques such as \textit{attention checks}, \textit{ground truths}, and \textit{interaction analysis} for quality assurance~\cite{kosara2010mechanical}. 

In this paper, we report the expectations, findings, and lessons learned from two ``failed'' Mechanical Turk experiments in which we aimed to evaluate a mixed-initiative visualization system. The research agenda was motivated by the need to improve data exploration for ``small'' but crowded data visualizations. In particular, for high-density data, visualizing every data point can lead to overplotting and information overload. Although, there are several methods for improving visual clutter such as filtering and sampling~\cite{liu_immens_2013}, these methods largely focus on ``big” data. Information overload can still occur in the small data settings, and naïve data reduction methods applied to a small samples can remove elements that are important to the user or exaggerate irrelevant points.

Our research aimed to manage overplotting by presenting a mixed-initiative information visualization system. The design uses a hidden Markov model algorithm, developed by Ottley et.~al~\cite{Ottley2019}, to capture and predict user attention. The visualization then responds by emphasizing the points that likely fits the user's interest. After conducting two trials of a large scale crowd-sourced experiment to study the effect of the system described in this paper, we found no evidence to support our hypotheses. Furthermore, our analysis revealed two significant and unexpected findings:
\begin{itemize}
    \item Participants just wanted to hover. We used clicking to trigger the visualization adaptation. However, the subjects in our online experiment overwhelmingly interacted with the visualization by hovering. 
    
    \item Open-ended tasks were not appropriate for our online studies. A vast majority of our online participants failed to provide the quality of feedback that we expected and that were produced by our in-person pilot studies. 
\end{itemize}

\section{Mixed-Initiative System Overview}\label{cpt:mixed_initiative} 
\begin{figure}[hb!]
    \centering
    \begin{subfigure}{0.4\textwidth}
        \centering
        \includegraphics[width=\textwidth]{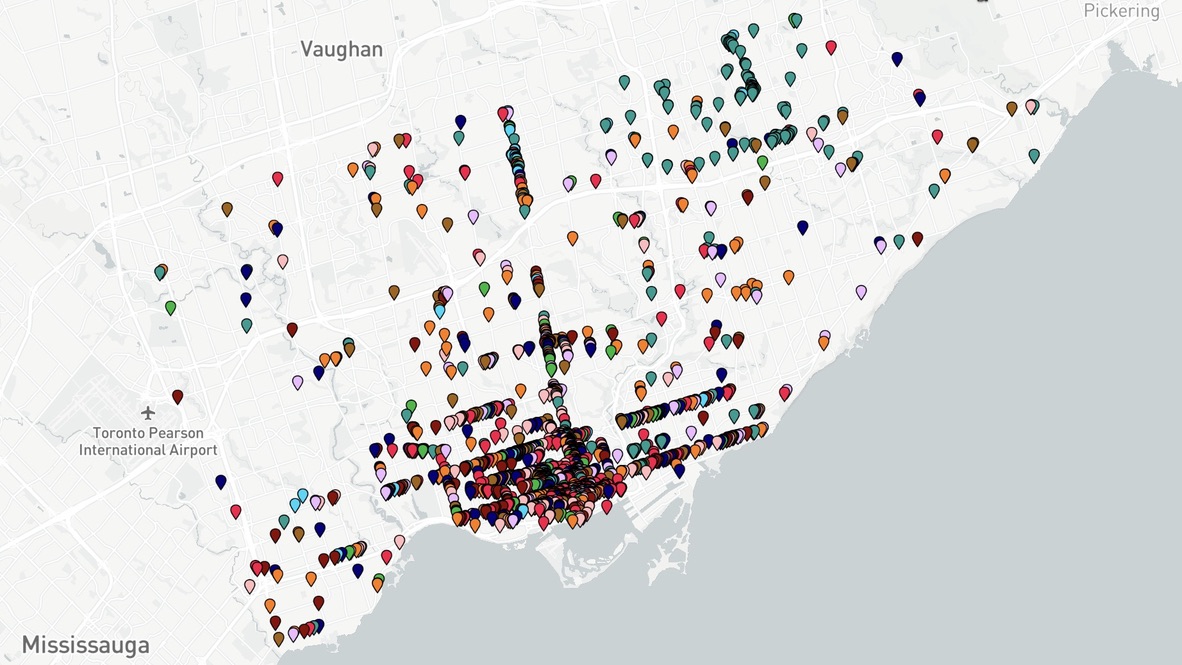}
    \end{subfigure}
    \begin{subfigure}{0.4\textwidth}
        \centering
        \includegraphics[width=\textwidth]{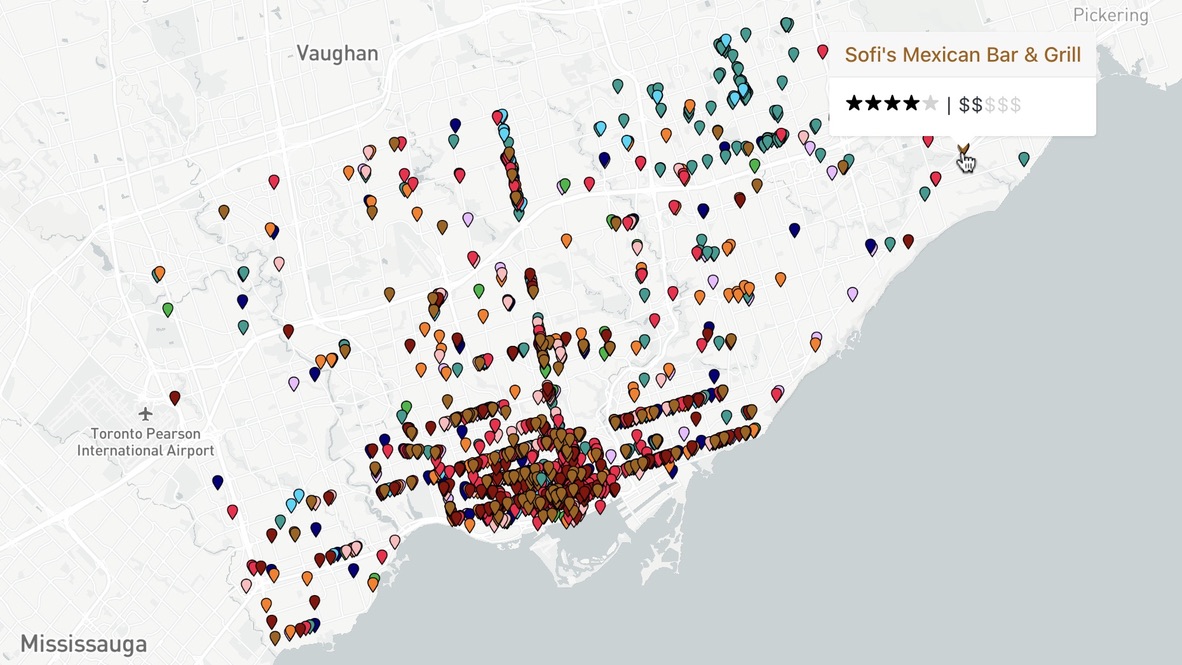}
    \end{subfigure}
    \caption{The changing z-order of the pins as the user clicks, with a particular interest in Mexican restaurants (coded in brown).}\label{fig:zorder}
\end{figure}

Our proposed mixed-initiative system was straightforward. We utilized the algorithm presented in Ottley et al.~\cite{Ottley2019} that captures and predicts users' attention, to promote potential points of interest to the user's interface. Consider for example, a visualization with large amounts of occlusion where datapoints overlap and partially or fully obscure each other. The user would typically use zooming to handle this occlusion, but this solution has its problems and limitations. When the visualization is completely zoomed out, many of the datapoints are partially or entirely occluded, making large trends hard to see. When the visualization is completely zoomed in, focal points can be seen in their entirety, but such a close view can cause the analyst to lose context~\cite{Jakobsen2013}. Our visualization system responds by adaptively re-drawing datapoints, bringing datapoints that the user is likely to be interested in to the foreground, and sending ``uninteresting'' datapoints to the background (as seen in Figure~\ref{fig:zorder}). In doing so, we hoped to create a system that allows for an informative overview, encourages easier exploration, and reduces the need for visual transformations.

\subsection{Interface Design}
The system's interface is displayed in Figure~\ref{fig:exp_main}. Our experiment used two datasets, Toronto restaurants~\cite{Yelp} and St.~Louis crimes~\cite{Crime}. We use pins the location of each data point on the map and color-coded then according to categories. For the Toronto dataset, the color of indicates the main cuisine for a given restaurant, and the color of the pins for the St.~Louis crime data indicates the type of crime. Details of points are provided in two forms: (1) a tooltip that appear on hovered, and (2) ``information cards'' are adding to the sidebar on click. The hover tooltip, as seen in segment 1 of Figure~\ref{fig:exp_main}, shows just a few details, such as the restaurant's name, rating, and price, or the crime's type and description. The information cards, on the other hand, show all the attributes of a given datapoint, as seen in segment 2 Figure~\ref{fig:exp_main}. The information cards also provide two more modes of interaction. The first is ``View'', which transitions the viewport to center and zoom in on the datapoint corresponding to that information card and temporarily enlarges the selected datapoint to bring the user's attention to the pin. The second is ``Delete'', which removes the card from the sidebar. This allows the user to keep a running list of datapoints of interest, and refer back to them on the map on demand. A legend at the bottom right of the screen, also serves as a filter, as seen in segment 3 of Figure~\ref{fig:exp_main}.

\section{Testing Our System}
We conducted two online experiments to determine the effect of a real-time adaptive system on user interactions and system usability. For each study, we recruited 200 participants via Amazon's Mechanical Turk. Our experiment used the two datasets introduced in the previous section and two conditions (responsive and unresponsive) to study the effects of adaptive systems, creating four groups:
\begin{enumerate}[noitemsep]
    \item Responsive Toronto, then Unresponsive St.~Louis
    \item Responsive St.~Louis, then Unresponsive Toronto
    \item Unresponsive Toronto, then Responsive St.~Louis
    \item Unresponsive St.~Louis, then Responsive Toronto
\end{enumerate}

If a user is on the responsive session of the experiment, the system adapts the visualization to the user's interests. In the unresponsive session, the system will not change the visualization at all as the user interacts.  

\subsection{Task Design}\label{sec:task_design}
Task design is critical to the success of an evaluation~\cite{munzner2009nested}. As a result, we carefully considered the evaluation of our system and explored a variety of task taxonomies (e.g., ~\cite{amar2005low} and \cite{yi2007toward}). Ultimately, we wanted to focus and exploratory data analysis. Specifically, we distinguish between \textit{bottom-up exploration} and \textit{top-down exploration}.  Bottom-up explorations ``are driven in reaction to the data"~\cite{alspaugh2018futzing} or ``may be triggered by salient visual cues"~\cite{Liu2014}. Top-down explorations, on the other hand, are based on a high-level goals or hypothesis~\cite{battle2019characterizing, Liu2014}. We settled on an open-ended task because we wanted to observe the users' instinctual behavior. We conducted a series of in-person pilot studies to determine the best phrasing for open-ended tasks prompts.

\subsection{Procedure and Data Collection}\label{sec:procedure}
At the start of the experiment, the participants were randomly assigned to one of the four groups. Inspired by a laboratory study on latency by Liu and Heer~\cite{Liu2014}, the participants were first asked to ``take some time to interact with the dataset in front of [them], exploring the data and gathering insights''. Once the participants felt that they were familiar with the dataset, they were given the opportunity to write down as many (or as few) insights as they would like. The participants were primed for interaction and insight gathering. Before the experiment started, examples of insights were shown to users (e.g., ``There are more kid-friendly coffee shops Downtown than there are Uptown''). The goal of this priming was to introduce the participants to the visualization and make explicit the idea of an ``insight'' without biasing the user during either segment of the experiment. In this way, users were free to perform exploratory data analysis without guidance or restriction, creating a general-purpose task that can demonstrate the flexibility of the adaptive system. Additionally, the reward structure of the experiment was designed to encourage this insight-gathering: participants were awarded \(\$1\) for participating, and \(\$0.50\) for every insight gathered. After recording all the insights that they found, the participants completed the System Usability Scale (SUS)~\cite{Brooke1996}, a widely-used, ``robust and versatile tool for usability professionals''~\cite{Bangor2008}, with an added comments section at the end of each survey for general comments from users. Upon completion of the survey, the participants continued onto the second condition/visualization which follows the same procedure as the first.

As the participants interacted with the system, we captured every mouse interaction: clicks, hovers, zooms, pans, views, deletes, and filter toggles. To separate intentional from unintentional hovers, we only recorded hovers with duration of at least 250 milliseconds. Additionally, we captured all insights, interaction time, and survey responses.

\section{What We Expected}
Before publishing our study on Mechanical Turk, there were some expectations that we put on our users. 

\begin{itemize}
    \item One of the main ways that users would interact with our system was through clicking on data points.
    \item The users would be able to provide useful insights about the data. 
    \item The responsive system would elicit fewer zooms and more insights.
\end{itemize}

Overall, we hoped to see the mixed-initiative system have a significant positive impact on the users' interactions and subjective feedback.

\section{The Reality of Study 1}
In our first study, hovering over a data point revealed a tooltip as shown in Figure~\ref{fig:exp_main}. Clicking on a point adds a card to the sidebar that shows a historical log of click interactions.

\begin{figure}[!h]
    \centering
    \includegraphics[width=\columnwidth]{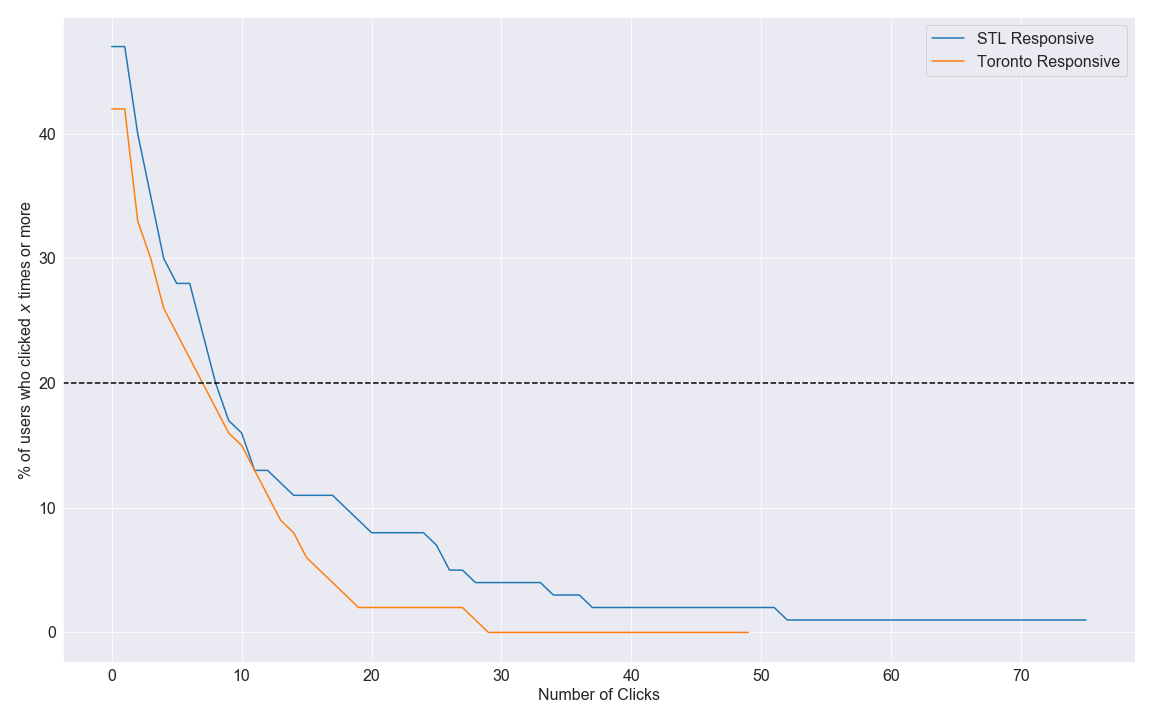}
    \caption{A graph from our first experiment showing how often users clicked on the visualization within a condition. As the number of clicks increases on the \(x\)-axis, the percentage of users who clicked at least \(x\) times decreases. The dotted line represents the users whose click interactions were within the top 20\% of their cohort. }\label{fig:user_clicks}
\end{figure}

\noindent Participants spent, on average, 27 minutes exploring the two datasets and recording their insights. 

\textbf{People wanted to hover.}
Unsurprisingly, we observed a large variance in the number of interactions. However, on average, participants clicked on 9 out of 2915 points on the Toronto map and 7 out of 1951 points in the St.\ Louis map. Figure~\ref{fig:user_clicks} plots the percentage of users by the number click performed during a session. Since, we used clicks as input to the machine learning algorithm that tracks and predicts future interactions, our responsive conditions were largely ineffective. We found no evidence that the adaptive system impacted analysis, and there was no indication that participants even noticed the adaptation. For further analysis, we narrowed the dataset to a group of users whose interactions met the expectations of the experimental design (41 participants remained). Table~\ref{tab:interactions} displays the average number of each type of interaction, spread across datasets and conditions.

\begin{table}[ht!]
    \centering
    \begin{tabularx}{\linewidth}{lcccc}
        \toprule
        {} & \multicolumn{2}{c}{\small Toronto} & \multicolumn{2}{c}{\small St.~Louis} \\
        \cmidrule(lr){2-3}
        \cmidrule(lr){4-5}
        {} & \small Responsive & \small Unresponsive & \small Responsive & \small Unresponsive \\
        \midrule
        Clicks   & 7.9      & 12.6      & 11.8      & 11.0         \\
        Hovers   & 52.0     & 41.0      & 46.4      & 38.6         \\
        Zooms    & 57.4     & 25.4      & 45.8      & 35.6         \\
        Insights & 5.3      & 4.8       & 5.5       & 6.2          \\
        \bottomrule
    \end{tabularx}
    \caption{Average values for interaction data frequency from the first trial. \label{tab:interactions}}
\end{table}

\textbf{Insights were shallow. }
We coded insights and categorized them as \textit{deep} and \textit{shallow} based on the amount of information that they contain. A \textit{shallow} insight is an observations attained only through minimal interactions. For example,

\begin{itemize}[noitemsep]
    \item [] \textbf{ID893}: ``There is a large concentration of Chinese restaurants in one area.''
    \item [] \textbf{ID950}: ``A majority of the crimes committed are theft-related.''
\end{itemize}
  A \textit{deep} insight requires building knowledge of the dataset through interactions. For example, 
  \begin{itemize}[noitemsep]
    \item [] \textbf{ID821}: ``There must be a Chinatown, or Asian-American populated area down Spadina Avenue between and including Dundas Street West, College Street, and Beverley Street.''

    \item [] \textbf{ID795}: ``There are incidences of larceny scattered all over the area. Most have to do with automobiles, but burglaries near the river tend to involve burglaries in businesses and buildings.''
\end{itemize}
  
Participants in our study entered a total of 779 insights, of which 747 were \textit{shallow} and 32 were classified as \textit{deep}. We believed that the lack of differences between the conditions were due to a flaw in the system design: the information given on-hover was nearly identical to the information given on-click. It is possible that this information parity gave no incentive for users to click on data points, other than to keep their information in a persistent state on the sidebar. This was particularly problematic in the context of our experiment that relied on user clicks to trigger the experimental condition.

\section{The Reality of Study 2}
Due to the flaws in the interface design and the overall null results, we rerun the experiment and address the design missteps causing the discrepancy between expectations and reality. We kept much of the experiment design from Experiment 2, but made two minor changes:
\begin{enumerate}[noitemsep]
    \item We removed the tooltip on hover. This meant that participants only saw details on click and clicking also triggered in the system adaptation. 
    \item We added additional guidance for formulating insights. In addition to the examples detailed in Section~\ref{sec:procedure}, the instruction dissuaded shallow insights by stating ``Obvious insights like `There are a lot of coffee shops' will not be rewarded the bonus.'' 
\end{enumerate}

\noindent
In this second study, participants spent an average of 29 minutes exploring the two datasets and recording their insights. 

\textbf{People did not want to click.} We observed a moderate increase in the number of clicks. On average, participants clicked on 40 out of 2915 points on the Toronto map and 23 out of 1951 points on the St. Louis map. However, we observed a pattern similar to Experiment 1. 110 out of 200 participants clicked on fewer than 5 data points. An analysis of the remaining 90 participants revealed inconclusive results (see Table ~\ref{tab:interactions2}).

\begin{table}[ht!]
    \centering
    \begin{tabularx}{\linewidth}{lcccc}
    \toprule
    {} & \multicolumn{2}{c}{\small Toronto} & \multicolumn{2}{c}{\small St.~Louis} \\
    \cmidrule(lr){2-3}
    \cmidrule(lr){4-5}
    {} & \small Responsive & \small Unresponsive & \small Responsive & \small Unresponsive \\
    \midrule
    Clicks   &       36.5 &         63.0 &       31.0 &         21.2 \\
    Hovers   &       50.8 &         67.6 &       48.1 &         41.2 \\
    Zooms    &       58.5 &         54.6 &       38.6 &         49.2 \\
    Insights &        4.8 &          5.3 &        5.6 &          4.7 \\
    \bottomrule
    \end{tabularx}
    \caption{Average values for interaction data frequency from the second trial.}
    \label{tab:interactions2}
\end{table}

\textbf{Insights were shallow again.}
Similar to the click finding, we observed an increase in the number of insights and moderate improvement in the quality of insights. Participants in the second study entered a total of 1045 insights, of which 996 were \textit{shallow} and 79 were classified as \textit{deep}.  

\section{Lessons Learned}
The finding that people may not interact with visualization in the way that we expect them to is not new. In the storytelling realm, Boy et al.~\cite{boy2015storytelling} found the participants in the web-based field experiments did not engage with visualization as expected. Reports from \textit{New York Times} suggest that users prefer scrolling as a means of interaction, which as led them to reconsider their investment in interactive visualization~\cite{tse2016why}. These are only anecdotal results, however, along with the findings of the ``failed'' user studies in this paper, they echo the sentiments of Lam~\cite{lam2008framework} who encourages designers to weigh the cost against potential gains of interaction.

Many researchers believe the ``the purpose of visualization is insight''~\cite{card1999readings}. In our studies, we opted for open ended tasks and captured insights in addition to quantitative measures. Although there is prior work that define~\cite{chang2009defining} and  characterize~\cite{saraiya2005insight} insights, insight-based evaluative methods, especially for online studies, are not clear. Many of the existing studies (e.g., \cite{liu_immens_2013} and \cite{saraiya2005insight}) captured insights in laboratory settings. In addition, the open-ended nature of our tasks made it difficult to the filter participants who were clicking through to get paid. This potentially highlights an important limitation of online studies and provides suggestive evidence that open-ended tasks may not be appropriate for this experiment setting. 

Not all of our findings were negative. The fact that users found the visualization well-designed, yet were overwhelmed by the size and density of the datasets, indicates that the chosen datasets worked well to induce a need for an assistive or mixed-initiative agent to help users make sense of the data. When designing this adaptive system, we were nervous that constantly updating the visualization would be disorienting to the user, especially given failed examples of adaptive systems like the notorious \textit{Microsoft Office Clippy}. However, the usability results for both responsive and non-responsive conditions were overwhelmingly positive and we saw no impact on usability for the responsive conditions. Overall, these lessons learned are motivating and we believe that we can move forward to create and evaluate a more rigorous and robust mixed-initiative system that actively supports the user during exploration.

\section{Future Directions}
The failures we discussed opens up next directions for our adaptive system. 
\begin{enumerate}[noitemsep]
    \item In both experiments, we saw that a majority of the users liked to hover to interact and gain insights. Like clicks, hovers could provide us a better understanding of the users' interests in real-time. A possible solution to improve the system would be to incorporate hovers along with clicks to the algorithm presented in Ottley et.~al~\cite{Ottley2019}. 
    \item Again, in both experiments, we had difficulty in obtaining insights that were deep and in good quality. It would be interesting to see if a mix of closed and open-ended tasks will help increase the quality of insights from Mechanical Turk users. The closed-ended tasks would be asked first to get the users familiar with interacting with the system. Then, the users would be free to explore and complete the open-ended tasks. 
\end{enumerate}

\section{Concluding Remarks}
We designed a mixed-initiative system and attempted to investigate the effect of the system on user interaction and system usability. It is tempting to say that results we found supports the alternative hypothesis and conclude that ``the adaptive system did nothing''. However, it is more accurate to say that we do not have enough evidence to reject the null hypothesis. Why is this a crucial distinction? Because it does not destroy any hope of an effective mixed-initiative system. It is important to note that these results should not dissuade researchers from further work on mixed-initiative systems like the one we designed. Although we were unsuccessful in achieving the data we expected, we found significant and unexpected findings from the users. When developing these systems, it is easy to assume that the users will interact the way that we want them to. Since we made an assumption that users would mainly show interest by clicking on the points, our system was unable to aide the users when their interactions did not meet our expectations. We have learned from our failures and hope that the VIS community can also learn from them too.

\acknowledgments{This work was supported in part by the National Science Foundation under Grant No. 1755734.}

\bibliographystyle{abbrv-doi}
\bibliography{failfest}

\end{document}